\documentclass{jetp}
\twocolumn
\begin{document}
\English
\title{Quasi-isotropic expansion for a two-fluid cosmological model containing radiation and stringy gas}
%\date{}
\setaffiliation1{L.D.Landau Institute for Theoretical Physics, Kosygin Str. 2,
119334 Moscow, Russia}
\setaffiliation2{Dipartimento di Fisica e Astronomia, Universit\`a di Bologna and INFN, Via Irnerio 46, 40126
Bologna, Italy}
\setauthor{I.~M.}{Khalatnikov}{1}
\email{khalat@itp.ac.ru}
\setauthor{A.~Yu.}{Kamenshchik}{12}
\email{kamenshchik@bo.infn.it}
\setauthor{A.~A.}{Starobinsky}{1}
\email{alstar@landau.ac.ru}

\abstract{
The quasi-isotropic expansion for a simple two-fluid cosmological model, including radiation and stringy gas is constructed.
The first non-trivial order expressions for the metric coefficients, energy densities and velocities are explicitly written down.
Their small and large time asymptotics are studied. It is found that the large time asymptotic for the anisotropic component 
of the metric coefficients grows faster than that of the isotropic (trace-proportional) component.}

\maketitle
\section{Introduction}
The 
quasi-isotropic solution of the Einstein equations near a cosmological 
singularity was found by Lifshitz and Khalatnikov \cite{Lif-Khal} for the 
Universe filled by radiation  with the equation of state $p = 
\frac{\varepsilon}{3}$ in the early 60th. 
In the paper \cite{we}, we presented the generalization of the 
quasi-isotropic solution of the Einstein equations near a cosmological 
singularity to the case of an arbitrary one-fluid 
cosmological model.
Then this solution was further generalized to the 
case of the Universe filled by two ideal barotropic fluids \cite{we-two}.

As is well known, modern cosmology deals with many very different 
types of matter. In comparison with the old standard model of the hot 
Universe (the Big Bang), the situation has been dramatically changed, 
first, with the development of inflationary cosmological models which 
contain an inflaton effective scalar field or/and other exotic types 
of matter as an important ingredient \cite{inflation,inflation1,inflation2,inflation3,inflation4,inflation5}, 
and second, with 
the understanding that the main part of the non-relativistic matter in 
the present Universe is non-baryonic -- cold dark matter (CDM). 
Furthermore, the appearance of brane and M-theory cosmological models~
\cite{brane,brane1} and the discovery of the cosmic acceleration~ 
\cite{cosmic,cosmic1} (see also \cite{varun} for a review) suggests that 
matter playing an essential role at different stages of cosmological 
evolution is multi-component generically, and these components may 
obey very different equations of state.  Moreover, the very notion of 
the equation of state appears to be not fundamental; it has only a 
limited range of validity as compared to a more fundamental 
field-theoretical description. From this general point of view, the 
generalization of the quasi-isotropic solution to the case of two 
ideal barotropic fluids with constant but different $p/{\varepsilon}$ 
ratios seems to be a natural and important next logical step. 

To explain the physical sense of the quasi-isotropic solution, let us 
remind that it represents the most generic spatially inhomogeneous 
generalization of the FRW space-time in which the space-time is locally
FRW-like near the cosmological singularity $t=0$ (in particular, its
Weyl tensor is much less than its Riemann tensor). On the other hand, 
generically it is very inhomogeneous globally and may have a very 
complicated spatial topology. As was shown in \cite{Lif-Khal1,we} (see 
also \cite{Deruelle,Deruelle1}), such a solution contains 3 arbitrary functions 
of space coordinates. From the FRW point of view, these 3 degrees of 
freedom represent the growing (non-decreasing in terms of metric 
perturbations) mode of adiabatic perturbations and the non-decreasing 
mode of gravitational waves (with two polarizations) in 
the case when deviations of a space-time metric from the FRW one are 
not small. So, the quasi-isotropic solution is not a generic solution 
of the Einstein equations with a barotropic fluid. Therefore, one 
should not expect this solution to arise in the course of generic 
gravitational collapse (in particular, inside a black hole event 
horizon). The generic solution near a space-like curvature singularity 
(for $p<\varepsilon$) has a completely different structure consisting 
of the infinite sequence of anisotropic vacuum Kasner-like eras with 
space-dependent Kasner exponents \cite {BLH,BLH1,Misner}.

For this reason, the quasi-isotropic solution had not attracted 
much interest for about twenty years. Its new life began after the 
development of successful inflationary models (i.e., with "graceful 
exit" from inflation) and the theory of generation of perturbations
during inflation, because it had immediately become clear that 
generically (without fine tuning of initial conditions) scalar metric
perturbations after the end of inflation remained small in a finite 
region of space which was much less than the whole causally connected 
space volume produced by inflation. It appears that the 
quasi-isotropic solution can be used for a global description of a 
part of space-time after inflation which belongs to "one 
post-inflationary universe". The latter is defined as a connected part 
of space-time where the hyper-surface $t=t_f({\bf r})$ describing the 
moment when inflation ends is space-like and, therefore, can be made 
the surface of constant (zero) synchronous time by a coordinate 
transformation. This directly follows from the derivation of 
perturbations generated during inflation given in \cite{Star82} (see 
Eq. (17) of that paper) which is valid in case of large perturbations, 
too. Thus, when used in this context, the quasi-isotropic solution 
represents an {\em intermediate} asymptotic regime during expansion of 
the Universe after inflation. The synchronous time $t$ appearing in it
is the proper time since the {\em end} of inflation, and the region 
of validity of the solution is from $t=0$ up to a moment in future 
when spatial gradients become important. For sufficiently large 
scales, the latter moment may be rather late, even of the order or
larger than the present age of the Universe. Note also  the analogue 
of the quasi-isotropic solution {\em before} the end of inflation is 
given by the generic quasi-de Sitter solution found in \cite{Star83}.
Both solutions can be smoothly matched across the hypersurface of 
the end of inflation.

A slightly different versions of the quasi-isotropic expansion was developed 
during last decades which are known under names of long-wave expansion or gradient expansion 
\cite{long-wave,long-wave1,long-wave2,long-wave3,long-wave4,long-wave5,long-wave6,long-wave7}. 

Originally the quasi-isotropic expansion was developed as a technique of generation of some kind of perturbative expansion 
in the vicinity of the cosmological singularity, where the cosmic time parameter served as a perturbative one. However,
the more general treatment of the quasi-isotropic expansion is possible if one notices that the next order of the 
quasi-isotropic expansion contain the higher orders of the spatial derivatives of the metric coefficients. 
Thus, it is possible to construct a natural generalization of the quasi-isotropic solution of the Einstein equations 
which would be valid not only in the vicinity of the cosmological singularity, but in the full time range.
In this case the simple algebraic equations, which one resolves to find the higher orders of the quasi-isotropic 
approximation in the vicinity of the singularity are substituted by differentially equations, where the time dependence 
of the metric can be rather complicated in contrast to the power-law behaviour of the coefficents of the original 
quasi-isotropic expansion. 

In the present paper we construct this expansion for a relativaly simple two-fluid cosmological model, containing 
radiation and cosmic string gas (see e.g. \cite{Kam-Khal}). Such a model has a technical advantage: the corresponding Friedmann equation is exactly solvable in terms of cosmic time and, hence, the cosmic time parameter is a natural one  for the quasi-isotropic  solution.
In the second section of the paper we explicitly construct the first non-trivial order expressions for the metric, 
energy densities and velocities of two fluids and write down their asymptotics at small and big values of the time parameter.
The last section contains some concluding remarks. In the Appendix we apply the developed formalism to the case of one-fluid cosmological model. In this case the solutions valid in the full time range coincide with those valid in the vicinity of singularity  \cite{we}.

\section{Calculation of the metric, energy densities and velocities in the first non-trivial order 
of the quasi-isotropic expansion}
We consider a model of the universe filled with two barotropic fluids:
the radiation satisfying the equation of state
\begin{equation}
p_R = \frac{1}{3} \varepsilon_R, 
\label{rad}
\end{equation}
and the stringy gas with the equation of state
\begin{equation}
p_S = -\frac{1}{3} \varepsilon_S.
\label{string}
\end{equation}
The solution of the Friedmann equation for such a flat universe filled with a mixture of these two fluids can be expressed 
as a function of the cosmic time $t$ as 
\begin{equation}
a(t) = \sqrt{At^2 + Bt},
\label{Friedmann}
\end{equation}
where $a(t)$ is the cosmological radius. Thus, we can take as a zero-level for the quasi-isotropic approximation 
the following spatial metric in the synchroneous reference system:
\begin{equation} 
ds^2 = dt^2 - \gamma_{\alpha\beta}dx^{\alpha}dx^{\beta},
\label{sinch}
\end{equation}
\begin{equation}
\gamma_{\alpha\beta} = a_{\alpha\beta}(x)(t + b(x)t^2),
\label{gamma0}
\end{equation}
where the Greek indices are spatial and $x$ stays also for spatial coordinates. As usual for two-fluid model 
the lowest quasi-isotropic approximation is given by two functions a symmetric tensor $a_{\alpha\beta}(x)$ and a scalar 
$b(x)$ \cite{we-two}. 
We shall look for the next approximation of the quasi-isotropic expansion, which will be proportional to the square of the spatial derivatives of these two functions:
\begin{equation} 
\gamma_{\alpha\beta} = a_{\alpha\beta}(x)(t + b(x)t^2) + c_{\alpha\beta}(x,t).
\label{gamma1}
\end{equation}
In what follows we omit the argument $x$ from the corresponding functions. We shall need the following expressions:
the inverse metric 
\begin{equation}
\gamma^{\alpha\beta} = \frac{a^{\alpha\beta}}{bt^2+t} - \frac{c^{\alpha\beta}}{(bt^2+t)^2}.
\label{inverse} 
\end{equation}
Here the matrix $a^{\alpha\beta}$ is defined by relation 
\begin{equation}
a^{\alpha\gamma}a_{\gamma\beta} = \delta^{\alpha}_{\beta}
\label{inverse1}
\end{equation}
and the indices in the metrix $c$ are raised and lowered by matrices $a^{\alpha\beta}$ and $a_{\alpha\beta}$.  However, the  
indices in the matrices $\gamma_{\alpha\beta}$ and in curvature the extrinsic curvature tensors they are raised and lowered 
by whole matrices $\gamma^{\alpha\beta}$ and $\gamma_{\alpha\beta}$.
The extrinsic curvature in the first approximation is 
\begin{equation}
K_{\alpha\beta} = (2bt + 1)a_{\alpha\beta} + \dot{c}_{\alpha\beta},
\label{external}
\end{equation}
where ``dot'' as usual means the differentiation with respect to the time parameter $t$.  
Then we have 
\begin{equation}
K_{\alpha}^{\beta} = \frac{(2bt + 1)\delta_{\alpha}^{\beta}}{bt^2 + t} + \frac{\dot{c}_{\alpha}^{\beta}}{bt^2 + t} 
- \frac{(2bt + 1)c_{\alpha}^{\beta}}{(bt^2 + t)^2},
\label{external1}
\end{equation}
\begin{equation}
K = K_{\alpha}^{\alpha} = \frac{3(2bt + 1)}{bt^2 + t} + \frac{\dot{c}}{bt^2 + t} - \frac{(2bt + 1) c}{(bt^2 + t)^2},
\label{external2}
\end{equation}
\begin{eqnarray}
&&\frac{\partial K_{\alpha}^{\beta}}{\partial t} = \frac{2b\delta_{\alpha}^{\beta}}{bt^2 + t} - 
\frac{(2bt + 1)^2\delta_{\alpha}^{\beta}}{(bt^2 + t)^2} \nonumber \\
&&+\frac{\ddot{c}_{\alpha}^{\beta}}{bt^2 + t} - \frac{2(2bt + 1) \dot{c}_{\alpha}^{\beta}}{(bt^2 + t)^2}\nonumber \\
&&-\frac{2bc_{\alpha}^{\beta}}{(bt^2 + t)^2} + \frac{2(2bt + 1)^2 c_{\alpha}^{\beta}}{(bt^2 + t)^3},
\label{external3}
\end{eqnarray}
\begin{eqnarray}
&&\frac{\partial K_{\alpha}^{\alpha}}{\partial t} = \frac{6b}{bt^2 + t} - 
\frac{3(2bt + 1)^2}{(bt^2 + t)^2} +\frac{\ddot{c}}{bt^2 + t} \nonumber \\
&&- \frac{2(2bt + 1) \dot{c}}{(bt^2 + t)^2} -\frac{2bc}{(bt^2 + t)^2} + 
\frac{2(2bt + 1)^2 c}{(bt^2 + t)^3},
\label{external4}
\end{eqnarray}
\begin{eqnarray}
&&K_{\alpha}^{\beta}K_{\beta}^{\alpha} = \frac{3(2bt + 1)^2}{(bt^2 + t)^2} + \frac{2(2bt + 1)\dot{c}}{(bt^2 + t)^2}\nonumber \\
&&-\frac{2(2bt + 1)^2c}{(bt^2 + t)^3}.
\label{external5} 
\end{eqnarray}
Now we can write down the expression for $R_0^0$ component of the Ricci tensor using the known formula \cite{LL,Lif-Khal}:
\begin{equation}
R_0^0 = -\frac12\frac{\partial K_{\alpha}^{\alpha}}{\partial t} - \frac14K_{\alpha}^{\beta}K_{\beta}^{\alpha}.
\label{R00}
\end{equation}
Thus,
\begin{eqnarray}
&&R_0^0 = \frac{3(2bt + 1)^2}{4(bt^2 + t)^2} - \frac{3b}{bt^2 + t} - \frac{\ddot{c}}{2(bt^2 + t)}\nonumber \\
&&+ \frac{(2bt + 1)\dot{c}}{2(bt^2 + t)^2} + \frac{bc}{(bt^2 + t)^2} - \frac{(2bt + 1)^2c}{2(bt^2 + t)^3}.
\label{R001}
\end{eqnarray}

The energy momentum tensor for a perfect fluid has the form
\begin{equation}
T_{i}^{j} = (\varepsilon + p)u_{i}u^{j} - p\delta_{i}^{j},
\label{energy-momentum}
\end{equation}
where $u_{i}$ is a four-velocity normalized as usual as 
\begin{equation}
u_i u^i = 1,
\label{velo} 
\end{equation}
or, in other words, as 
\begin{equation}
u_0^2 - \gamma^{\alpha\beta}u_{\alpha}u_{\beta} = 1.
\label{velo1}
\end{equation}
In our approximation the spatial components of the four-velocities will be proportional to the spatial gradients of the 
scalar function $b$:
\begin{equation}
u_{\alpha} = v b_{,\alpha}.
\label{velo2}
\end{equation}
The trace of the energy momentum tensor is equal to 
\begin{equation}
T = T_i^i = \varepsilon - 3p.
\label{en-mom-tr}
\end{equation}
For our mixture of two fluids with the equations of state (\ref{rad}) and (\ref{string}) one has
\begin{equation}
T = 2\varepsilon_S.
\label{en-mom-tr1}
\end{equation}
We shall choose the Newton constant in such a way that the Einstein equations look as 
\begin{equation}
R_{i}^j = T_i^j - \frac{1}{2}\delta_i^j T.
\label{Einstein}
\end{equation}
Then, the temporal-temporal component of the system of the Einstein equations is 
\begin{equation}
R_{0}^0 = T_0^0 - \frac{1}{2} T.
\label{Einstein1}
\end{equation}
Using expressions (\ref{energy-momentum}-\ref{en-mom-tr}) one gets 
\begin{eqnarray}
&&T_0^0 - \frac{1}{2} T = \varepsilon_R + \frac43 \varepsilon_R v_R^2 b_{,\alpha}b_{,\beta}\gamma^{\alpha\beta}\nonumber \\
&&+ \frac23 \varepsilon_S v_S^2 b_{,\alpha}b_{,\beta}\gamma^{\alpha\beta}.
\label{en-mom1}
\end{eqnarray}
Substituting Eqs. (\ref{en-mom1}) and (\ref{R001}) into Eq. (\ref{Einstein1}) and taking into consideration only 
the zero-order terms, we get 
\begin{equation}
\varepsilon_R^{(0)} = \frac{3}{4(bt^2 + t)^2}.
\label{energyR0}
\end{equation}
The spatial-spatial components of the Ricci tensor are given by the formula \cite{LL}:
\begin{equation}
R_{\alpha}^{\beta} = -P_{\alpha}^{\beta} - \frac12 \frac{\partial K_{\alpha}^{\beta}}{\partial t} - 
\frac14 K_{\gamma}^{\gamma}K_{\alpha}^{\beta},
\label{Ricci}
\end{equation}
where $P^{\alpha}_{\beta}$ is the Ricci tensor constructed from the spatial metric $\gamma_{\mu\nu}$.
The Ricci scalar is 
\begin{equation}
R = R_0^0 + R_{\alpha}^{\alpha} = -P - \frac{\partial K_{\alpha}^{\alpha}}{\partial t} - \frac14 (K_{\alpha}^{\alpha})^2 - 
\frac14 K_{\alpha}^{\beta}K_{\beta}^{\alpha}.
\label{Ricci-scalar} 
\end{equation}
Now, using the contracted Einstein equation
\begin{equation}
R = -T 
\label{Einstein-con}
\end{equation}
together with Eq. (\ref{en-mom-tr1}) and taking into account only the zero-order contributions, one oobtains 
\begin{equation}
\varepsilon_S^{(0)} = \frac{3b}{bt^2 + t}.
\label{energyS0}
\end{equation}

We shall need also the expression for the mixed, spatial-temporal components of the Ricci tensor:
\begin{equation}
R_{\alpha}^0 = \frac12 (K_{\alpha|\beta}^{\beta} - K_{\beta|\alpha}^{\beta}),
\label{Ricci-mixed}
\end{equation}
where the covariant derivative is taken by using the Christoffel symbols constructed from the metric $\gamma_{\mu\nu}$.
In the first non-vanishing order
\begin{equation}
R_{\alpha}^0 = -\frac{t^2}{(bt^2 + t)^2} b_{,\alpha}.
\label{Ricci-mixed1}
\end{equation}
The mixed component of the energy-momentum tensor is in the same order
\begin{equation}
T_{\alpha}^0 = \left(\frac43 \varepsilon_R^{(0)}v_R + \frac23 \varepsilon_S^{(0)}v_S\right)b_{,\alpha} 
\label{en-mom-mixed}
\end{equation}
Comparing Eqs. (\ref{Ricci-mixed1}) and (\ref{en-mom-mixed}) and using expressions (\ref{energyR0}) and (\ref{energyS0}) 
we get
\begin{equation}
v_R + 2b(bt^2+t)v_S = -t^2.
\label{Einstein-mixed}
\end{equation}

Eq. (\ref{Einstein-mixed}) is not sufficient to find the velocities $v_R$ and $v_S$ and we should use the energy-momentum 
tensor conservation laws. We suppose that our fluids do not interact and hence the corresponding energy momentum tensors shoud be conserved separately. The spatial component of the energy-momentum tensor conservation law can be written down as
\begin{equation}
T_{\alpha , 0}^0 + \frac12 K_{\beta}^{\beta} T_{\alpha}^0 - p_{,\alpha} = 0.
\label{conserv}
\end{equation}
This equation implies that  radiation and  string gas velocity functions, respectively, satisfy the following 
first-order differential equations:
\begin{equation}
\dot{v}_R -\frac16 K_{\beta}^{\beta} v_R -\frac{1}{4\varepsilon_R}\frac{d\varepsilon_R}{db} = 0,
\label{velReq}
\end{equation}
\begin{equation}
\dot{v}_S +\frac16 K_{\beta}^{\beta} v_S +\frac{1}{2\varepsilon_S}\frac{d\varepsilon_S}{db} = 0.
\label{velSeq}
\end{equation}
Solving these equations we obtain
\begin{equation}
v_R = -\frac{\sqrt{bt^2+t}}{2b^{3/2}}{\rm  Arccosh}(2bt+1) + \frac{t}{b},
\label{velR}
\end{equation}
\begin{equation}
v_S = -\frac{1}{2b^2} + \frac{1}{4b^{5/2}\sqrt{bt^2+t}}{\rm  Arccosh}(2bt+1).
\label{velS}
\end{equation}
It is easy to check that these velocities satisfy Eq. (\ref{Einstein-mixed}). 

We shall need also the $0$ component of the energy-momentum conservation law, which looks like
\begin{eqnarray}
&&T_{0,0}^0 + T_{0,\alpha}^{\alpha} - \Gamma_{0 \alpha}^{\beta}T_{\beta}^{\alpha} + \Gamma_{\alpha 0}^0 T_0^0\nonumber \\
&&+ \Gamma_{\alpha\beta}^{\alpha}T_0^{\beta} = 0.
\label{conserv1}
\end{eqnarray}
Up to first order the relevent quantities are written as 
\begin{equation}
T_0^0 = \varepsilon^{(0)} + \varepsilon^{(1)} + \frac{(\varepsilon + p)^{(0)} v^2 b_{,\alpha}b_{,}^{\alpha}}{bt^2+t},
\label{T001}
\end{equation}
\begin{equation}
T_0^{\alpha} = -\frac{(\varepsilon + p)^{(0)} v b_{,}^{\alpha}}{bt^2+t},
\label{T0alpha1}
\end{equation}
\begin{equation}
T_{\beta}^{\alpha} = -\delta_{\beta}^{\alpha} p^{(0)} -\delta_{\beta}^{\alpha} p^{(1)} 
- \frac{(\varepsilon + p)^{(0)} v^2 
b_{,\beta}b_{,}^{\alpha}}{bt^2+t},
\label{Talphabeta}
\end{equation}
\begin{equation}
\Gamma_{0\alpha}^{\beta} = \frac12 K_{\alpha}^{(0) \beta} + \frac12 K_{\alpha}^{(1) \beta}.
\label{Chris-ext} 
\end{equation}
Using Eqs. (\ref{conserv1})--(\ref{Chris-ext}) one can find the following equations for the first corrections to 
the enrgy densities $\varepsilon^{(1)}$:
\begin{eqnarray}
&&\dot{\varepsilon}^{(1)} + \frac{d}{dt} \left(\frac{(\varepsilon + p)^{(0)} v^2 
b_{,\alpha}b_{,}^{\alpha}}{bt^2+t}\right)
\nonumber \\
&&+\frac12 K^{(0)} \left(\varepsilon^{(1)} + p^{(1)} + \frac{4(\varepsilon + p)^{(0)} v^2 b_{,\alpha}b_{,}^{\alpha}}{3(bt^2+t)}\right)\nonumber \\
&&+\frac12 K^{(1)}(\varepsilon + p)^{(0)} - \left(\frac{(\varepsilon + p)^{(0)} 
vb_{,}^{\alpha}}{bt^2+t}\right)_{|\alpha} = 0.
\label{dif-for-en}
\end{eqnarray}
The equations (\ref{dif-for-en}) could be explicitly integrated for both the energy densities $\varepsilon^{(1)}_R$ and $\varepsilon^{(1)}_S$,
expressing their relation to other unknown  quantity: the trace of the first correction to the metric $c$:
\begin{equation}
\varepsilon_R^{(1)} = -\frac{c}{2(bt^2+t)^3} -\frac43 \frac{\varepsilon_R^{(0)}v_R^2 b_{,\alpha}b_{,}^{\alpha}}{bt^2+t}+\bar{\varepsilon}_R^{(1)},
\label{en-rad-0}
\end{equation}
where
\begin{wide}
\begin{eqnarray}
&&\bar{\varepsilon}_R^{(1)} =
-\left(\frac14{\rm Arccosh}^2(2bt+1) - 
\ln(bt+1)\right)\frac{b_{|\alpha}^{\alpha}}{b^2(bt^2+t)^2}\nonumber \\
&&+\frac{1}{b^3(bt^2+t)^2}\left(-2\ln(1+bt) + \frac{3b^2t^2}{b(bt^2+t)} \right. \nonumber \\
&&\left. +{\rm Arccosh}^2(2bt+1) -\frac{5bt{\rm Arccosh}(2bt+1)}{2\sqrt{b}\sqrt{bt^2+t}}\right)b_{,\alpha}b_{,}^{\alpha},
\label{en-rad-1}
\end{eqnarray}
\begin{equation}
\varepsilon_S^{(1)} = -\frac{bc}{(bt^2+t)^2}-\frac23 \frac{\varepsilon_S^{(0)}v_S^2 b_{,\alpha}b_{,}^{\alpha}}{bt^2+t}
+\bar{\varepsilon}_S^{(1)},
\label{en-st-0}
\end{equation}
where
\begin{eqnarray}
&&\bar{\varepsilon}_S^{(1)} = 
+\frac{b_{|\alpha}^{\alpha}}{b(bt^2+t)}\left(2\ln(bt+1) +2 -\frac{(2bt+1){\rm Arccosh}(2bt+1)}{b^{1/2}(bt^2+t)^{1/2}} \right) 
\nonumber \\
&&+\frac{b_{,\alpha}b_{,}^{\alpha}}{4b^2(bt^2+t)}\left(\frac{{\rm Arccosh}^2(2bt+1)}{4b^2(bt^2+t)^2} 
+\frac{(4bt-1){\rm Arccosh}(2bt+1)}{b^{3/2}(bt^2+t)^{3/2}} \right. \nonumber \\
&&\left. +\frac{6-48bt-b^2t^2}{6b(bt^2+t)}\right).
\label{en-st-1} 
\end{eqnarray}
Now combining the Einstein equations in such a way to exclude of them the term including the second time derivatives of $c$ 
we have the equation 
\begin{eqnarray}
&&2\varepsilon_R^{(1)} + 2\varepsilon_S^{(1)} - P + \frac{1}{4}K_{\alpha}^{\beta}K_{\beta}^{\alpha} 
-\frac14(K_{\alpha}^{\alpha})^2 
\nonumber \\
&& +\frac83 \frac{\varepsilon_R^{(0)}v_R^2 b_{,\alpha}b_{,}^{\alpha}}{bt^2+t} 
+\frac43 \frac{\varepsilon_S^{(0)}v_S^2 b_{,\alpha}b_{,}^{\alpha}}{bt^2+t} = 0.
\label{eq-for-c}
\end{eqnarray}
We shall need also the following expression for the spatial curvature scalar
\begin{equation}
P = \frac{\bar{P}}{bt^2+t} - \frac{2t^2}{(bt^2+t)^2}b_{|\alpha}^{\alpha} + \frac{3t^4}{2(bt^2+t)^3}b_{,\alpha}b_{,}^{\alpha},
\label{space-cur}
\end{equation}
where $\bar{P}$ denotes the spatial curvature scalar constructed from the metric $a_{\alpha\beta}$.
 
Then 
\begin{equation}
c = (2bt+1)\times 
\int dt \frac{(bt^2+t)^2}{(2bt+1)^2}\left(-P + 2\bar{\varepsilon}_R^{(1)} 
 + 2\bar{\varepsilon}_S^{(1)} \right).
\label{Eq-for-c}
\end{equation}

The expression for the trace of the correction to the metric is 
\begin{eqnarray}
&&c = -\bar{P} \frac{t^2}{2} \nonumber \\
&&+\frac{b_{|\alpha}{^\alpha}}{b^3}\times\left(b^2t^2 - bt + 2bt(1+bt)\ln(bt+1) + (2bt+1)\ln(2bt+1)\right. \nonumber \\
&&\left.-(2bt+1)\sqrt{b^2t^2+bt}{\rm Arccosh}(2bt+1)+\frac14{\rm Arccosh}^2(2bt+1)\right) \nonumber \\
&&+\frac{(2bt+1)b_{,\alpha}b_{,}^{\alpha}}{b^4}\times\left(\frac{-19b^2t^2-85bt}{24(2bt+1)}+\frac{4(bt+1)\ln(bt+1)}{2bt+1}\right.
\nonumber \\
&&-\frac{9\ln(bt+1)}{2}+\frac{61\ln(2bt+1)}{48}-\frac{3{\rm Arccosh}^2(2bt+1)}{4(2bt+1)}\nonumber \\
&&+\frac{2{\rm Arccosh}(2bt+1)\sqrt{b^2t^2+bt}}{2bt+1}+\frac18{\rm Arccosh}^2(2bt+1)\ln\frac{bt}{bt+1}\nonumber \\
&&-{\rm Arccosh}(2bt+1)\left({\rm Li}_2(e^{-{\rm Arccosh}(2bt+1)})-\frac14{\rm Li}_2(e^{-2{\rm Arccosh}(2bt+1)})\right)\nonumber \\
&&\left.-\left({\rm Li}_3(e^{-{\rm Arccosh}(2bt+1)})-\frac18{\rm Li}_3(e^{-2{\rm Arccosh}(2bt+1)})\right)+\frac78\zeta_R(3)\right),
\label{trace-complete}
\end{eqnarray}
\end{wide}
where $\zeta_R(n)$ is the Riemann zeta function, while ${\rm Li}_n(x)$ is the polylogarithm function defined as 
\begin{equation}
{\rm Li}_n(x) = \sum_{k=1}^{\infty}\frac{x^k}{k^n}.
\label{polylog}
\end{equation}
Now we would like to obtain the expressions for the obtained quantities at small  values of the time parameter $t$. 
We shall need the following formulae:
\begin{eqnarray}
&&{\rm Arccosh}(2bt+1) = \ln (1 + 2bt +2\sqrt{bt + b^2t^2}) \nonumber \\
&&= 2\sqrt{bt} -\frac13 (bt)^{3/2} + \frac{3}{20} (bt)^{5/2} + o((bt)^{5/2}),
\label{arccosh-as}
\end{eqnarray}
\begin{equation}
{\rm Arccosh}^2(2bt+1) = 4bt -\frac43 (bt)^2 + \frac{32}{45} (bt)^3 + o((bt)^3).
\label{arccosh-as2}
\end{equation}

At the limit of small values of time parameter $t$ (which coincides with the limit of small $b$) one has for $c$ the following expression:
\begin{equation}
c = -\bar{P} \frac{t^2}{2} - \frac{11}{9}b_{|\alpha}^{\alpha} t^3 -\frac{91 b_{,\alpha}b_{,}^{\alpha}}{101b}t^3.
\label{c-asymp}
\end{equation}
At large values of the time parameter one has 
\begin{equation}
c = -\bar{P} \frac{t^2}{2} + \frac{b_{|\alpha}^{\alpha}t^2}{b} - \frac{19 b_{,\alpha}b_{,}^{\alpha} t^2}{24b^2}.
\label{c-asymp-big}
\end{equation}
Thus, at large values of time $t$ the corrections to the metric represents some tensor depending only on special coordinates, 
multiplied by $t^2$.

Now, substituting into the formulae  (\ref{en-rad-0})
 and (\ref{en-st-0}) the expressions (\ref{trace-complete}), (\ref{en-rad-1}), (\ref{en-st-1}) and the velocities  (\ref{velR}) and 
(\ref{velS}) we obtain
\begin{wide}
\begin{eqnarray}
&&\varepsilon_R^{(1)} = \frac{\bar{P}t^2}{4(bt^2+t)^3}\nonumber \\
&&+\frac{b_{| \alpha}^{\alpha}}{2(b^2t^2+bt)^3}(-b^2t^2 + bt - bt(1+bt)\ln(1+bt) -(1+2bt)\ln(1+2bt)\nonumber \\
&&\left.+((1+2bt)\sqrt{b^2t^2+bt}{\rm Arccosh}(2bt+1)-\frac14(2b^2t^2 + 2bt + 1){\rm Arccosh}^2(2bt+1)\right)\nonumber \\
&&+\frac{b_{,\alpha}b_{,}^{\alpha}}{b^4(bt^2+t)^3}\left(\frac{115b^2t^2 + 85bt}{48} + \left(\frac52bt+\frac14\right)\ln(1+bt)\right.
\nonumber \\
&&-\frac{192b^2t^2+318bt+61}{96}\ln(1+2bt)-\frac52bt\sqrt{b^2t^2+bt}{\rm Arccosh}(2bt+1)\nonumber \\
&&-\frac{2bt+1}{16}{\rm Arccosh}^2(2bt+1)\ln\frac{bt}{1+bt} + \frac{3(2b^2t^2+2bt+1)}{8} {\rm Arccosh}^2(2bt+1) \nonumber \\
&&+\frac{2bt+1}{2}\left({\rm Arccosh}(2bt+1)({\rm Li}_2(e^{-{\rm Arccosh}(2bt+1)})-\frac14 {\rm Li}_2(e^{-2{\rm Arccosh}(2bt+1)})\right)\nonumber \\
&&\left. +\left({\rm Li}_3(e^{-{\rm Arccosh}(2bt+1)})-\frac18 {\rm Li}_3(e^{-2{\rm Arccosh}(2bt+1)})\right)-\frac78\zeta_R(3)\right)
\label{en-rad-final}
\end{eqnarray}
and 
\begin{eqnarray}
&&\varepsilon_S^{(1)} = \frac{\bar{P}bt^2}{2(bt^2+t)^2} \nonumber \\
&&+\frac{b_{| \alpha}^{\alpha}}{b^2(bt^2+t)^2}\left(b^2t^2+3bt-(1+2bt)\ln(1+2bt)-\frac14{\rm Arccosh}^2(2bt+1)\right)\nonumber \\
&&+\frac{b_{, \alpha}b_{,}^{\alpha}}{b^3(bt^2+t)^2}\left(\frac{18b^2t^2+37bt-6}{24}+\left(5bt+\frac12\right)\ln(1+bt)\right.\nonumber \\
&&-\frac{61(1+2bt)\ln(1+2bt)}{48}-\frac{1+2bt}{8}{\rm Arccosh}(2bt+1)\ln\frac{bt}{1+bt}\nonumber \\
&&+\frac{1-4bt-8b^2t^2}{4\sqrt{b^2t^2+bt}}{\rm Arccosh}(2bt+1)+\frac{12b^2t^2+12bt-1}{16b(bt^2+t)}{\rm Arccosh}^2(2bt+1)\nonumber\\
&&+(1+2bt){\rm Arccosh}(2bt+1)\left({\rm Li}_2(e^{-{\rm Arccosh}(2bt+1)})-\frac14{\rm Li}_2(e^{-2{\rm Arccosh}(2bt+1)})\right)\nonumber \\
&&+(1+2bt)\left({\rm Li}_3(e^{-{\rm Arccosh}(2bt+1)})-\frac18{\rm Li}_3(e^{-2{\rm Arccosh}(2bt+1)})\right)\nonumber \\
&&\left.-\frac{7(1+2bt)}{8}\zeta_R(3)\right).
\label{en-st-final}
\end{eqnarray}
\end{wide}
At the small time limit the energy densities (\ref{en-rad-final}) and (\ref{en-st-final}) look like
\begin{equation}
\varepsilon_R^{(1)} = \frac{\bar{P}}{4t} +\frac49b_{| \alpha}^{\alpha} + \frac{91}{202}\frac{b_{, \alpha}b_{,}^{\alpha}}{b},
\label{en-rad-final1}
\end{equation}
\begin{equation}
\varepsilon_S^{(1)} = \frac{\bar{P}b}{2} - \frac23 b_{| \alpha}^{\alpha} - \frac{103}{90}\frac{b_{, \alpha}b_{,}^{\alpha}}{b}.
\label{en-st-final1}
\end{equation}
 
When the time parmeter $t \rightarrow \infty$ one has
\begin{equation}
\varepsilon_R^{(1)} = \frac{1}{4b^3t^4}\left(\bar{P} - \frac{b_{| \alpha}^{\alpha} \ln^2 bt}{b} + \frac{3b_{,\alpha}b_{,}^{\alpha}\ln^2 bt}{b^2}\right),
\label{en-rad-final2}
\end{equation}
\begin{equation}
\varepsilon_S^{(1)} = \frac{1}{4bt^2}\left(2\bar{P} + \frac{2(1-4\ln2)b_{| \alpha}^{\alpha}}{b} - \frac{3b_{,\alpha}b_{,}^{\alpha}}{b^2}\right).
\label{en-st-final2}
\end{equation}

To find the traceless part of the first correction to the metric 
\begin{equation}
\tilde{c}_{\alpha\beta} \equiv c_{\alpha\beta} - \frac13 a_{\alpha\beta}c 
\label{tilde-def}
\end{equation}

we use the Einstein equation for the spatial components of the Ricci and the energy-momentum tensor, taking their traceless parts:
\begin{equation}
\tilde{R}_{\alpha}^{\beta} = \tilde{T}_{\alpha}^{\beta}.
\label{Einstein-traceless}
\end{equation}
From Eqs. (\ref{Ricci}) and (\ref{external2})  one finds that the traceless part of the spatial-spatial components of the Ricci tensor is 
\begin{equation}
\tilde{R}_{\alpha}^{\beta} = -\tilde{P}_{\alpha}^{\beta} - \frac12 \frac{\partial \tilde{K}_{\alpha}^{\beta}}{\partial t} - 
\frac{3(2bt + 1)}{4(bt^2 + t)}\tilde{K}_{\alpha}^{\beta},
\label{Ricci-traceless}
\end{equation}
while from Eq. (\ref{Talphabeta}) one finds that 
\begin{equation}
\tilde{T}_{\alpha}^{\beta} = - \frac{(\varepsilon + p)_R^{(0)} v_R^2 + (\varepsilon + p)_S^{(0)} v_S^2)
\left(b_{,\alpha}b_{,}^{\beta}-\frac{1}{3}\delta_{\alpha}^{\beta}b_{,\gamma}b_{,}^{\gamma}\right)}{bt^2+t}.
\label{Talphabeta-traceless} 
\end{equation}
We shall need also the expression for the spatial Ricci tensor 
\begin{equation}
P_{\alpha}^{\beta} = \frac{\bar{P}_{\alpha}^{\beta}}{bt^2+t} - \frac{2t^2}{(bt^2+t)^2}b_{;\alpha}^{\beta} + \frac{3t^4}{2(bt^2+t)^3}b_{,\alpha}b_{,}^{\beta}.
\label{space-cur-Ricci}
\end{equation}
Now combining Eqs. (\ref{Einstein-traceless}) and (\ref{Ricci-traceless}) we obtain the following equation 
for the traceless part of the extrinsic curvature:
\begin{equation}
(bt^2+t)^{-3/2}\frac{\partial (\tilde{K}_{\alpha}^{\beta}(bt^2+t)^{3/2})}{\partial t} = -2\tilde{P}_{\alpha}^{\beta} - 
2\tilde{T}_{\alpha}^{\beta}, 
\label{traceless-extr-eq}
\end{equation}
which immediately gives
\begin{equation}
\tilde{K}_{\alpha}^{\beta} = -2(bt^2+t)^{-3/2}\int (bt^2+t)^{3/2})(\tilde{P}_{\alpha}^{\beta}+
\tilde{T}_{\alpha}^{\beta}) dt.
\label{integral-ext}
\end{equation}
Substituting into Eq. (\ref{integral-ext}) expressions (\ref{Talphabeta-traceless}), (\ref{space-cur-Ricci}),  
(\ref{energyR0}), (\ref{energyS0}), (\ref{velR}), (\ref{velS}) one finds the following expression for the 
traceless part of the extrinsic curvature:
\begin{wide}
\begin{eqnarray}
&&\tilde{K}_{\alpha}^{\beta} = -\left(\frac{2bt+1}{2b(bt^2+t)}-\frac{{\rm Arccosh}(2bt+1)}{4b^{3/2}(bt^2+t)^{3/2}}\right)\tilde{\bar{P}}_{\alpha}^{\beta}\nonumber \\
&&+\frac{1}{2b^{5/2}(bt^2+t)^{3/2}}(2b^{1/2}\sqrt{bt^2+t}(2bt-3)+3{\rm Arccosh}(2bt+1))\left(b_{;\alpha}^{\beta}-\frac13 \delta_{\alpha}^{\beta}b_{;\gamma}^{\gamma}\right)\nonumber \\
&&+\frac{1}{b^{7/2}(bt^2+t)^{3/2}}
\times\left(\frac{bt(29+15bt-6b^2t^2)}{4\sqrt{b^2t^2+bt}}-\frac{21}{8}{\rm Arccosh}(2bt+1) \right. \nonumber \\
&&\left. +\frac16 {\rm Arccosh}^3(2bt+1) -
\frac{(2bt+1){\rm Arccosh}^2(2bt+1)}{2\sqrt{b^2t^2+bt}}\right)
\left(b_{,\alpha}b_{,}^{\beta}-\frac13\delta_{\alpha}^{\beta}b_{,\gamma}b_{,}^{\gamma}\right).
\label{K-traceless}
\end{eqnarray}

The expression for the traceless part of the correction to the metric $\tilde{c}_{\alpha\beta}$ can be found as 
\begin{equation}
\tilde{c}_{\alpha}^{\beta} = (bt^2+t) \int \tilde{K}_{\alpha}^{\beta} dt.
\label{integral-for-c}
\end{equation}
Finally for the traceless part of $c_{\alpha\beta}$ we obtain

\begin{eqnarray}
&&\tilde{c}_{\alpha\beta} = \left(\frac{bt^2+t}{b}-\frac{(2bt+1)\sqrt{b^2t^2+bt}}{2b^2}{\rm Arccosh}(2bt+1)\right)\tilde{\bar{P}}_{\alpha\beta}\nonumber \\
&&+\frac{bt^2+t}{b^2}\left(8\ln(1+bt)+ 6 -\frac{3(2bt+1){\rm Arccosh}(2bt+1)}{\sqrt{b^2t^2+bt}}\right)
\left(b_{;\alpha\beta}-\frac{1}{3}a_{\alpha\beta}b_{;\gamma}^{\gamma}\right)\nonumber \\
&&+\frac{bt^2+t}{b^3}\left(-\frac{2bt}{1+bt}-16\ln(1+bt)+\zeta_R(3)-{\rm Li}_3(e^{-2{\rm Arccosh}(2bt+1)})\right.\nonumber \\
&&-2{\rm Arccosh}^2(2bt+1){\rm Li}_2(e^{-2{\rm Arccosh}(2bt+1)})-\frac43{\rm Arccosh}^3(2bt+1) \nonumber \\
&&-\frac{(2bt+1){\rm Arccosh}^3(2bt+1)}{3\sqrt{b^2t^2+bt}}+{\rm Arccosh}^2(2bt+1)\ln(bt) + 4{\rm Arccosh}^2(2bt+1)\ln 2
\nonumber\\  
&&+ {\rm Arccosh}^2(2bt+1)\ln(1+bt)
+\frac{{\rm Arccosh}^2(2bt+1)}{2(b^2t^2+bt)}\nonumber \\
&&\left.+\frac{29(2bt+1){\rm Arccosh}(2bt+1)}{4\sqrt{b^2t^2+bt}}-\frac{33}{2}\right)\times \left(b_{,\alpha}b_{,\beta} - \frac13 a_{\alpha\beta}b_{,\gamma}b_{,}^{\gamma}\right).
\label{traceless}
\end{eqnarray} 
At small $t$ the expression (\ref{traceless}) behaves as 
\begin{equation}
\tilde{c}_{\alpha\beta} = -\frac43 t^2\tilde{\bar{P}}_{\alpha\beta} + \frac45 t^3 \left(b_{;\alpha\beta}-\frac{1}{3}a_{\alpha\beta}b_{;\gamma}^{\gamma}\right) 
+\frac{4}{45} \frac{t^3}{b} \left(b_{,\alpha}b_{,\beta} - \frac13 a_{\alpha\beta}b_{,\gamma}b_{,}^{\gamma}\right).
\label{traceless1}  
\end{equation}
At large values of $t$ it looks like 
\begin{equation}
\tilde{c}_{\alpha\beta} = t^2\ln bt \left(-\tilde{\bar{P}}_{\alpha\beta} + \frac2b\left(b_{;\alpha\beta}-\frac{1}{3}a_{\alpha\beta}b_{;\gamma}^{\gamma}\right)
-\frac{3}{2b^2}\left(b_{,\alpha}b_{,\beta} - \frac13 a_{\alpha\beta}b_{,\gamma}b_{,}^{\gamma}\right)\right).
\label{traceless2} 
\end{equation}
Thus, the large-time asymptotic behaviour of the traceless ``anisotropic '' part of the metric is very different from 
the asymptotic behaviour of the trace of the metric $c$ (\ref{c-asymp-big}). The anisotropic part of the metric grows faster 
( by logarithm $\ln bt$). 
\end{wide}
To understand better such a behaviour of the anisotropic part of the metric, let us remember that at large values of $t$ 
the string gas dominate radiation. One can make transition from the two-fluid case to one fluid-one, where only the string gas is present, considering the limit
\begin{eqnarray}
&&b \rightarrow \infty, \nonumber \\
&&b a_{\alpha\beta} = const.
\label{limit-string}
\end{eqnarray}
Substituting (\ref{limit-string})  into the expressions (\ref{trace-complete}) and (\ref{traceless}) one sees that while the limiting 
value of the trace part of the metric $c$ is regular, the coefficient $c_{\alpha\beta}$ diverges as $\ln bt$. 
It is quite natural because we know that first correction to the traceless part of the metric in the quasi-isotropic expansion 
diverges for the barotropic fluid with $w = -\frac13$ (see \cite{we} and the appendix).   

\section{Concluding remarks}
We have calculated explicitly the  first non-trivial order expressions for the metric (\ref{trace-complete}), (\ref{traceless}) 
and for energy-densities (\ref{en-rad-final}), (\ref{en-st-final}) and velocities  (\ref{velR}), (\ref{velS}) of two fluids.
We have written down also their asymptoric expressions for small and large values of the cosmic time parameter. As it was easily 
predictable, in the case of small $t$ the structure of the solution is determined only by radiation component and coincides 
with that found in the original paper \cite{Lif-Khal}. However, the large time behaviour of the metric coefficients reveals an unusual feature: the anisotropic part of the metric (\ref{traceless2}) grows essentially faster than the isotropic one 
(\ref{traceless1}) and the relation between these two speeds of growth behaves as $\ln bt$. It seems that his effect is connected 
with two-fluid character of the model, which had been considered in this paper. 
The other reason of this behaviour is connected with the fact that the quasi-isotropic expansion for the string gas has a singular character (see \cite{we} and the Appendix of the present paper). 
However, it is not clear if the effect of behaviour different form power-law  
is present in two-fluid models with other equations of state. To answer this question it is necessary to develop the formalism 
of building of the quasi-isotropic expansion valid for the full tange of time for arbitrary two-fluid models which is technically 
much more complicated.   

This work was partially supported by the RFBR grant No 17-02-01008.

%\appendix
\section*{Appendix}
\setcounter{section}{1}
In this Appendix we consider the quasi-isotropic expansion for one-fluid cosmological model, which is valid for the full time range.
The method of calculation is the same as in Sec. 2, but in one-fluid case they are much simpler. We shall see that  the form of  the expression 
for the first correction to the metric coefficients valid for the full time range coincides this that valid in the vicinity of the singularity \cite{we}, obtained  by the method first proposed in \cite{Lif-Khal}. 

We consider the universe with the fluid with the equation of state $p = w\varepsilon$. The spatial metric 
now is 
\begin{equation}
\gamma_{\alpha\beta} = a_{\alpha\beta}t^{\kappa} + c_{\alpha\beta},
\label{metricA}
\end{equation}
where
\begin{equation}
\kappa = \frac{4}{3(1+w)}.
\label{kappa}
\end{equation}
Inverse metric is
\begin{equation}
\gamma^{\alpha\beta} = a^{\alpha\beta}t^{-\kappa} - c^{\alpha\beta}t^{-2\kappa}.
\label{inverseA}
\end{equation}
Then we have the following formulae for the extrinsic curvature:
\begin{equation}
K_{\alpha\beta} = a_{\alpha\beta} \kappa t^{\kappa -1} + \dot{c}_{\alpha\beta},
\label{extrinsicA}
\end{equation}
\begin{equation}
K_{\alpha}^{\beta} = \delta_{\alpha}^{\beta} \kappa + \dot{c}_{\alpha}^{\beta}t^{-\kappa}-c_{\alpha}^{\beta}\kappa t^{-\kappa-1},
\label{extrinsicA1} 
\end{equation}
\begin{equation}
K = \frac{3\kappa}{t} + \dot{c}t^{-\kappa}-c\kappa t^{-\kappa-1},
\label{extrinsicA2}
\end{equation}
\begin{equation}
\frac{\partial K}{\partial t} = -\frac{3\kappa}{t^2} + \ddot{c}t^{-\kappa} - 2\dot{c}\kappa t^{-\kappa-1} + c\kappa(\kappa+1) t^{-\kappa-2},
\label{extrinsicA3}
\end{equation}
\begin{equation}
K_{\alpha}^{\beta}K_{\beta}^{\alpha}  = \frac{3\kappa^2}{t^2} + 2\dot{c}\kappa t^{-\kappa-1} - 2c\kappa^2t^{-\kappa-2}.
\label{extrinsicA4}
\end{equation}
Substituting formulae (\ref{extrinsicA3}), (\ref{extrinsicA4}) into (\ref{R00}) we have
\begin{equation}
R_{0}^{0} = \frac{3\kappa(2-\kappa)}{4t^2} - \frac{\ddot{c}t^{-\kappa}}{2} + \frac{\dot{c} \kappa t^{-\kappa-1}}{2} - \frac{c\kappa t^{-\kappa-2}}{2}.
\label{R00A}
\end{equation}
Using now the Einstein equation (\ref{Einstein1}) in the lowest order of the approximation we obtain for the energy density of the fluid under consideration 
\begin{equation}
\varepsilon^{(0)} = \frac{4}{3(1+w)^2t^2}.
\label{energy0A} 
\end{equation}
Using the 0 component of the energy-momentum conservation law (\ref{conserv1}) we can find the relation between 
the first correction to the energy density $\varepsilon^{(1)}$ and the trace of the first correction to the metric $c$:
\begin{equation}
\varepsilon^{(1)} = -\frac{c\kappa t^{-\kappa-2}}{2}.
\label{energyA1}
\end{equation}
Now, using the expression for the scalar curvature $R$ (\ref{Ricci-scalar}) and the  00 component of the Einstein equation in the form $R_0^0 - \frac12 R = T_0^0$ for in the fisrt quasi-isotropic order the following equation:
\begin{equation}
\frac{P}{2} + \frac14 K^{(0)}K^{(1)} - \frac18 (K_{\alpha}^{\beta}K_{\beta}^{\alpha})^{(1)} = \varepsilon^{(0)}.
\label{dif-eqA} 
\end{equation}
Combining (\ref{dif-eqA}) and (\ref{energyA1}) we obtain the following differential equation for $c$:
\begin{equation}
\dot{c} = \frac{c (\kappa -1)}{t} - \frac{\bar{P} t}{\kappa}.
\label{dif-eqA1}
\end{equation}
Integrating (\ref{dif-eqA1}) we obtain
\begin{equation}
c = \frac{\bar{P}t^2}{\kappa(\kappa-3)} = -\frac{9(1+w)^2\bar{P}t^2}{4(5+9w)}.
\label{cA}
\end{equation}

Now to find the traceless part of the first correction to the metric $\tilde{c}_{\alpha\beta}$ we shall use traceless 
part of the spatial-spatial component of the Einstein equations, which in the case of one fluid and in the first order approximation has a particularly simple form:
\begin{equation}
\tilde{R}_{\alpha}^{\beta} = -\tilde{P}_{\alpha}^{\beta} - \frac12\frac{\partial \tilde{K}_{\alpha}^{\beta}}{\partial t} -\frac{3\kappa}{4t}\tilde{K}_{\alpha}^{\beta} = 0.
\label{tracelessA}
\end{equation}
(Notice that the traceless part of the extrinsic curvature $\tilde{K}_{\alpha}^{\beta}$ does not have zero-order terms).
The equation (\ref{tracelessA}) can be rewritten as 
\begin{equation}
\frac{\partial \tilde{K}_{\alpha}^{\beta}}{\partial t} + \frac{3\kappa}{2t}\tilde{K}_{\alpha}^{\beta} = 
-2\tilde{\bar{P}}_{\alpha}^{\beta} t^{-\kappa}.
\label{tracelessA1}
\end{equation}
Integrating (\ref{tracelessA1}) one obtains
\begin{equation}
\tilde{K}_{\alpha}^{\beta} = -\frac{4}{\kappa+2}\tilde{\bar{P}}_{\alpha}^{\beta} t^{-\kappa+1}
\label{tracelessA2} 
\end{equation}
Using relation
\begin{equation}
\tilde{c}_{\alpha}^{\beta} = t^{\kappa}\int \tilde{K}_{\alpha}^{\beta} dt,
\label{relationA} 
\end{equation}
we come to 
\begin{equation}
\tilde{c}_{\alpha}^{\beta} = \frac{4}{\kappa^2-4}\tilde{\bar{P}}_{\alpha}^{\beta} t^2 = 
-\frac{9(1+w)^2}{(3w+1)(3w+5)}\tilde{\bar{P}}_{\alpha}^{\beta} t^2.
\label{tracelessA3}
\end{equation}

One can see that the results (\ref{cA}) and (\ref{tracelessA3}) valid in the full range of time coincide with 
those valid in the vicinity of the initial cosmological singularity ($t = 0$) \cite{we} obtained by the algebraic method 
\cite{Lif-Khal}. 
The general expression for the first correction to the metric for one-fluid case is given in the formula (37) in \cite{we}.
The metric $b_{\alpha\beta}$ in \cite{we} corresponds to $c_{\alpha\beta}$ in the present paper, while for the equation of 
state parameter the symbol $k$ is used instead of $w$.  
The formula (37) contains a misprint: in front of the second term in the brackets in the right-kand side of this equation 
should stay the factor $1/4$. At first glance the first correction to the metric (37) contains a pole at $3k+1 = 0$, however
calculating the trace of this metric, one sees that this pole is cancelled and is present  only in its  anisotropic part. 

Thus, for the case of the universe filled with the  string gas $w = -\frac13$ the quasi-isotropic expansion does not 
work because the expression for $\tilde{c}_{\alpha\beta}$ becomes singular. As we have seen before the quasi-isotropic 
expansion for the universe  filled with the mixture of string gas and radiation does work, but at large values of the time 
parameter $t$, when the influence of the string gas becomes dominant,  the metric coefficient $\tilde{c}_{\alpha\beta}$ 
 grows rapidly as $t^2\ln bt$ (\ref{traceless2}). 

In the conclusion let us consider a special case whan the metric $a_{\alpha\beta}$ has a conformally flat form:
\begin{equation}
a_{\alpha\beta} = e^{\rho(x)}\delta_{\alpha\beta}.
\label{flat}
\end{equation}
In this case the spatiall Ricci tensor is 
\begin{equation}
\bar{P}_{\alpha\beta} = \frac14(\rho_{,\alpha}\rho_{\beta} - 2 \rho_{,\alpha\beta}) - \frac14\delta_{\alpha\beta}(\rho_{,}^{\mu}
\rho_{,\mu} + 2 \rho_{,\mu}^{\mu})
\label{flat1}
\end{equation}
or 
\begin{equation}
\bar{P}_{\alpha}^{\beta} = \frac14(\rho_{,\alpha}\rho_{,}^{\beta} - 2 \rho_{,\alpha}^{\beta}) -\frac14\delta_{\alpha}^{|beta}
(\rho_{,}^{\mu}\rho_{,}^{\mu} + 2\rho_{,\mu}^{\mu}).
\label{flat2}
\end{equation}
Correspondingly
\begin{equation}
\bar{P} = -2\rho_{,\mu}^{\mu} -\frac12\rho_{,\mu}^{\mu}
\label{flat3}
\end{equation}
and the traceless part of the Ricci tensor is 
\begin{equation}
\tilde{\bar{P}}_{\alpha}^{\beta} = \frac14 (\rho_{,\alpha}\rho_{,}^{\beta} - 2\rho_{,\alpha}^{\beta}) 
+ \frac{1}{12} \delta_{\alpha}^{\beta}(2\rho_{,\mu}^{\mu} - \rho_{,\mu}\rho_{,}^{\mu}).
\label{flat4}
\end{equation}
If 
\begin{equation}
\rho = A_{\mu\nu}x^{\mu}x^{\nu}
\label{Gauss}
\end{equation}
then 
\begin{equation}
\bar{P} = -4A_{\mu}^{\mu} - 2A_{\mu\nu}A_{\alpha}^{\mu}x^{\nu}x^{\alpha}
\label{Gauss1}
\end{equation}
and 
\begin{equation}
\tilde{\bar{P}}_{\alpha}^{\beta} = \frac13 x^{\gamma}x^{\nu}(3A_{\alpha\gamma}A_{\nu}^{\beta} - \delta_{\alpha}^{\beta} A_{\mu\gamma}
A_{\nu}^{\mu}).
\label{Gauss2}
\end{equation}

Thus, it is easy to see that if the metric in the lowest order of the quasi-isotropic expansion has the Gaussian form 
determined by Eqs. (\ref{flat}) and (\ref{Gauss}) already its first correction $c_{\alpha\beta}$ determined by the 
cruvature tensors (\ref{Gauss1}) and (\ref{Gauss2}) has non-Gaussian form due to the presence of the quadratic 
in $x^{\mu}$ terms in front of the Gaussian exponential.

\end{document}